\documentclass{article}
\usepackage{graphicx}
\usepackage{amsfonts}
\usepackage{theorem}
\theorembodyfont{\rmfamily}
\newtheorem{Lem}{Lemma}[section]

\newtheorem{The}[Lem]{Theorem}
\newtheorem{Prop}[Lem]{Proposition}
\newtheorem{Cor}[Lem]{Corollary}

\newcommand{\qed}{\hbox{\rule{6pt}{6pt}}}

\begin{document}
\title{A precise estimation of the computational complexity in Shor's factoring algorithm}
\author{K.Kuriyama$^1$\footnote{E-mail:kuriyama@yamaguchi-u.ac.jp},
S.Sano$^1$\footnote{E-mail:sano@max.apsci.yamaguchi-u.ac.jp} and S. Furuichi$^2$\footnote{E-mail:furuichi@ed.yama.tus.ac.jp} 
 \\
$^1${\small Department of Applied Science, Faculty of Engineering,} \\
{\small Yamaguchi University,Tokiwadai 2-16-1, Ube City, 755-0811, Japan}\\
$^2${\small Department of Electronics and Computer Science,}\\
{\small Tokyo University of Science, Onoda City, Yamaguchi, 756-0884, Japan}}
\date{}
\maketitle
{\bf Abstract.} 
A precise estimation of the computational complexity in Shor's factoring algorithm under the condition 
that the large integer we want to factorize is composed by the product of finite prime numbers, 
is derived by the results related to number theory.  
\vspace{3mm}


\vspace{3mm}


\section{Introduction}
P.W.Shor proposed the algorithm to solve the factorization into prime factors within polynomial time by using the quantum
computer \cite{Shor1,Shor2}. See also \cite{MCA}, on the recent advances of the Shor's factoring algorithm.
In his paper, the number of the computational steps was estimated.
Through this paper, we consider the case that we factorize $n$ into finite prime numbers $p_i, (i=1,\cdots,k)$.
We derive the precise estimation of the number of the computational steps, and then
we show our estimation is more precise than Shor's original one.
Shor showed it is required at least $N$ times to be performed correctly factorization algorithm with the probability
$1-\varepsilon$ or higher than it for any $\varepsilon >0$, as follows
\begin{equation}
N \geq \frac{\log(1/\varepsilon)}{\alpha \beta (1 - 1/2^{k-1})} (\log_2 n)^2,
\end{equation}
where $\alpha,\beta$ are independent constant numbers with respect to $n = {p_1}^{e_1} {p_2}^{e_2} \dots {p_k}^{e_k}$.
In this paper, we improve the Shor's estimation of $N$ in the sense that our estimation is more precise than Shor's one. 
By putting $p_i - 1 = 2^{\tau_i} \sigma_i \,$ with odd numbers $\sigma_i$ and $\tau_i \geq 1,(i=1,2,\dots,k)$,
$\tau' = \min(\tau_1, \dots, \tau_k)$, $\tilde \tau = \sum^k_{i=1} \tau_i$, we have for any $\varepsilon >0$, 
\begin{equation}
N \geq \frac{\log(1/\varepsilon)}{\alpha \beta \left( 1 - \frac{1}{2^k - 1} 
\frac{2^k - 2 + 2^{k \tau'}}{2^{\tilde \tau}} \right) } \left( \log_2 n \right)^2.
\end{equation}
This paper is organaized as follows: In section 2, we will review the estimation by Shor to compare with ours in the final section.
In section 3, we show some results derived from number theory. In section 4, we show a new estimation of the computational 
complexity in Shor's factoring algorithm in the case that arbitrary integer $n$ is factorized into finite prime numbers $p_i, (i=1,\cdots,k)$,
applying the results obtained in the previous section. Finally in section 5, we discuss the relation between our estimation and Shor's one.

\section{An original estimation by Shor}
As mentioned in the above, we review the original estimation by Shor 
in the case that arbitrary integer $n$ is factorized into finite prime numbers $p_i,(i=1,\cdots,k)$.
However, for simplicity, we review the Shor's factoring algorithm in the case that arbitrary integer $n$ is factorized into two prime numbers $p$ and $q$ in the following manner.
\begin{itemize}
\item[$1^{\circ}$] Choose a number $a$ from the set $\{ 1,\dots,n-1 \}$.
\item[$2^{\circ}$] Calculate $\gcd(a,n)$.
\item[$3^{\circ}$] If $\gcd(a,n)=1$, then go to the next step. Otherwise, go to the step $8^{\circ}$.
\item[$4^{\circ}$] Calculate the order $r$ of the number $a$ with respect to $\bmod \,\, n$. (This calculations depend on the quantum computations.)
\item[$5^{\circ}$] If $r$ is even number, then go to the next step. Otherwise, go to the step $8^{\circ}$.
\item[$6^{\circ}$] Calculate $p' = \gcd(a^{r/2} + 1, n)$ and $q' = \gcd(a^{r/2} - 1, n)$.
\item[$7^{\circ}$] If either $p'$ or $q'$ is equal to $n$, then the next step. Otherwise, these number $p'$ and $q'$ are
the prime number we seek. 
\item[$8^{\circ}$] Go to the step $1^{\circ}$, and choose another $a$.
\end{itemize}

We denote the probability which succeeds in the factorization at the first try through the above algorithm by $P_S$.
We need at least $N$ times to find the prime number under the condition that $P_S \geq 1-\varepsilon$:
\begin{equation} \label{eq:N}
  N \geq \log(1 / \varepsilon) / P_S .
\end{equation}
In order to evaluate the probability $P_S$, we consider the following events:
\begin{itemize}
\item $A_a$ : The event which can be obtained $a$ satisfying $a<n$ and $\gcd(a,n) = 1$.
\item $A_r$ : The event which can be obtained the true order $r$ by quantum computations. 
\item $A_e$ : The event which the order $r$ becomes even number.
\item $A_f$ : The event which $p'$ and $q'$ become prime numbers $p$ and $q$ we seek.
\end{itemize}
By using the above notations, the probability $P_S$ can be represented by
\begin{equation}\label{eq:PS}
  P_S \geq P(A_e \cap A_f \mid A_a \cap A_r) P(A_a \cap A_r)
= P(A_e \cap A_f \mid A_a \cap A_r) P(A_a) P(A_r) .
\end{equation}
Therefore we evaluate three probability $P(A_r),P(A_a)$ and $P(A_e \cap A_f \mid A_a \cap A_r)$.
The event $A_r$  is equivalent to the event which obtains $d$ satisfying $d < r$ and $\gcd(d,r) = 1$ 
so that we have $P(A_r) = \varphi(r) / r$ by the use of Euler's function $\varphi$.
Since it is known that for Euler constant $\gamma$
\begin{equation}
  \liminf_{r \to \infty} \frac{\varphi(r) \log \log r}{r} = e^{-\gamma},
\end{equation}
we then  have for sufficient large $r$, 
\begin{equation}  \label{eq:PAr}
  P(A_r) = \frac{\varphi(r)}{r} \geq \frac{e^{-\gamma}}{\log \log r} \geq \frac{e^{-\gamma}}{\log r}
  \geq \frac{e^{-\gamma}}{\log n} = \frac{e^{-\gamma} \log_2 e}{\log_2 n} = \frac{\alpha}{\log_2 n},
\end{equation}
where $\alpha$ is independent constant number with respect to $n$.
Since $P(A_a) = \varphi(n) / (n-1)$, we similarly have 
\begin{equation}
  P(A_a) \geq \frac{\beta}{\log_2 n} \label{eq:PAa}
\end{equation}
where $\beta$ is independent constant number with respect to $n$.

Finally we note on the probability $P(A_e \cap A_f \mid A_a \cap A_r)$. 
In general, it is known that \cite{jozsa,NC} for $n = {p_1}^{e_1} {p_2}^{e_2} \dots {p_k}^{e_k}$
with different prime numbers $p_i (i=1,\cdots ,k)$, the order $r$ given by quantum computer is even number and 
\begin{equation}
  P(A_e \cap A_f \mid A_a \cap A_r) \geq 1 - \frac{1}{2^{k-1}}. \label{eq:PAef}
\end{equation}
(In the section 4, we precisely estimate this probability.)
Now we can estimate the computational complexity of Shor's algorithm.
Substituting Eq.(\ref{eq:PAa}),Eq.(\ref{eq:PAr}) and Eq.(\ref{eq:PAef}) into 
Eq.(\ref{eq:PS}), the probability $P_S$ is computed by
\begin{equation}
  P_S \geq \left( 1 - \frac{1}{2^{k-1}} \right) \frac{\alpha \beta}{(\log_2 n)^2}. \label{eq:PS2}
\end{equation}
From Eq.(\ref{eq:N}), we also have for any $\varepsilon >0$,
\begin{equation}
N \geq \frac{\log(1/\varepsilon)}{\alpha \beta (1 - 1/2^{k-1})} (\log_2 n)^2. \label{eq:N2}
\end{equation}

\section{Lemmas in number theory}

For the prime number $p$, we denote the field $  {\bf Z} / p {\bf Z}$ which all elements are invertible, by $({\bf Z} / p {\bf Z})^{\times}$. 
It is well known \cite{hw} that the following relation holds
\begin{equation}
  \left| \{a \in ({\bf Z} / p {\bf Z})^{\times} ; r_p = d \} \right| = \varphi(d) \label{eq:order},
\end{equation}
where $d \mid p-1$ and $r_p$ represents the order of $a$ with respect to $\bmod \,\, p$. Then we have the following two lemmas.

\begin{Lem}
For the number of the elements of $({\bf Z} / p {\bf Z})^{\times}$, we may write $p-1=2^{\tau}\sigma$ for odd number $\sigma$ and $\tau \geq 1$. Then we have 
\begin{eqnarray}  
  && \left| \{a \in ({\bf Z} / p {\bf Z})^{\times} ; r_p : odd \} \right| = \sigma \label{eq:p}\\
  && \left| \{a \in ({\bf Z} / p {\bf Z})^{\times} ; r_p = 2^t s \,(s:odd) \} \right| = 2^{t-1} \sigma ,\label{eq:pt}
\end{eqnarray}
where $t$ is a fixed number $(t=1,2,\cdots , \tau)$.
\end{Lem}
{\bf (Proof)}
For $r_p = 2^t s$ with odd number $s$ and  $t \geq 0$, the following equivalent relation holds
\begin{equation}
  r_p:odd, \,r_p \mid p-1 \Longleftrightarrow r_p \mid \sigma .
\end{equation}
Since $r_p \mid p-1 = 2^t s \mid 2^{\tau} \sigma$, we have $t \leq \tau, \,s \mid \sigma$, and then $t=0$ by the fact that $r_p = 2^t s$ is odd number.
Therefore we have $r_p = s$, and then we have $r_p \mid \sigma$ by $s \mid \sigma$. Conversely, if $r_p \mid \sigma$, then $r_p$
is the divisor of $\sigma$. Thus $r_p$ is odd number. Moreover, if $r_p \mid \sigma$, then $ r_p \mid p-1$, since $p-1 = 2^{\tau} \sigma$.
Thus we have 
\begin{eqnarray*}
  \left| \{a \in ({\bf Z} / p {\bf Z})^{\times} ; r_p : odd \} \right|
  &=& \sum_{r_p \mid p-1, \, r_p:odd} \varphi(r_p) \\
  &=& \sum_{r_p \mid \sigma} \varphi(r_p) \\
  &=& \sigma
\end{eqnarray*}

In the case of $r_p = 2^t s$, it holds the following equivalent relation
\begin{equation}
  r_p \mid p-1 \Longleftrightarrow s \mid \sigma .
\end{equation}
Indeed,
by assumption $1 \leq t \leq \tau$, if $2^t s \mid 2^{\tau} \sigma$, then we have $s \mid \sigma$.
Conversely $s \mid \sigma$ implies $s \mid 2^{\tau} \sigma$. Therefore for a fixed number $t$ in $(1 \leq t \leq \tau)$, we have
\begin{eqnarray*}
  \left| \{a \in ({\bf Z} / p {\bf Z})^{\times} ; r_p = 2^t s \,(s:odd) \} \right|
  &=& \sum_{r_p \mid p-1, \,r_p = 2^t s} \varphi(r_p) \\
  &=& \sum_{s \mid \sigma} \varphi(2^t s) \\
  &=& \sum_{s \mid \sigma} \varphi(2^t) \varphi(s) \\
  &=& \varphi(2^t) \sum_{s \mid \sigma} \varphi(s) \\
  &=& 2^t \left( 1 - \frac{1}{2} \right) \sigma \\
  &=& 2^{t-1} \sigma
\end{eqnarray*}
\hfill \qed

\begin{Lem}
For $n = {p_1}^{e_1} \dots {p_k}^{e_k} $ with prime numbers $p_i, (i=1,\cdots,k) $, we may write 
$p_i - 1 = 2^{\tau_i}$ with odd numbers  $\sigma_i$ and $\tau_i \geq 1$.
We denote the order of $a$ with respect to $\bmod \,\,n$ and $\bmod \,\,p_i$ by $r$ and $r_{p_i} = 2^{t_{p_i}} s_{p_i}$, respectively.
Then we have
\begin{eqnarray}
  && \left| \{a \in ({\bf Z} / n {\bf Z})^{\times} ; r : odd \} \right| = \prod_{i=1}^k \sigma_{p_i},
  \label{eq:pq} \\
  && \left| \{a \in ({\bf Z} / n {\bf Z})^{\times} ; t_{p_1} =\cdots = t_{p_k} = l \} \right| 
  = 2^{2(l-1)} \prod_{i=1}^k \sigma_{p_i}, \label{eq:pql}
\end{eqnarray}
where $l$ is a fixed number such that $1 \leq l \leq \min(\tau_{p_1}, \cdots, \tau_{p_k})$.
\end{Lem}
{\bf (Proof)}
From Chinese Remainder Theorem, we have $({\bf Z} / n {\bf Z})^{\times} \cong ({\bf Z} / p_1 {\bf Z})^{\times} \oplus\cdots \oplus ({\bf Z} / p_k {\bf Z})^{\times}$.
Then for $a \in ({\bf Z} / n {\bf Z})^{\times}$, we have
\begin{eqnarray}
  r &=& \textrm{lcm} \{ r_{p_1},\cdots, r_{p_k} \}, \nonumber \\
  r: odd & \Longleftrightarrow & r_{p_1},\cdots, r_{p_k} : odd, \nonumber \\
  \left| ({\bf Z} / n {\bf Z})^{\times} \right| &=&  \left| ({\bf Z} / p_1 {\bf Z})^{\times} \right| \cdots \left| ({\bf Z} / p_k {\bf Z})^{\times} \right|. \label{eq:crt}
\end{eqnarray}
Thus we have
\begin{eqnarray*}
   \left| \{a \in ({\bf Z} / n {\bf Z})^{\times} ; r : odd \} \right| 
  &=& \left| \{a \in ({\bf Z} / n {\bf Z})^{\times} ; r_{p_1},\cdots, r_{p_k} : odd \} \right| \\
  &=& \left| \{a \in ({\bf Z} / p_1 {\bf Z})^{\times} ; r_{p_1} : odd \} \right|\cdots
     \left| \{a \in ({\bf Z} / p_k {\bf Z})^{\times} ; r_{p_k} : odd \} \right| \\
  &=& \prod_{i=1}^k \sigma_{p_i}.
\end{eqnarray*}
Moreover, for a fixed $l$, we have
\begin{eqnarray*}
   \left| \{a \in ({\bf Z} / n {\bf Z})^{\times} ; t_{p_1} =\cdots = t_{p_k} = l \} \right| 
  &=& \left| \{a \in ({\bf Z} / n {\bf Z})^{\times} ; t_{p_1} = l \} \right|\cdots
     \left| \{a \in ({\bf Z} / n {\bf Z})^{\times} ; t_{p_k} = l \} \right| \\
  &=&  2^{2(l-1)} \prod_{i=1}^k \sigma_{p_i}
\end{eqnarray*}
\hfill \qed

Thanks to the above two lemmas, we have the following theorem.

\begin{The}
Define $\tau' \equiv \min(\tau_{p_1},\cdots, \tau_{p_k})$ and $\tilde \tau  \sum^k_{i=1} \tau_{p_i}$. Then we have
\begin{equation}
  \left| \{a \in ({\bf Z} / n {\bf Z} )^{\times} ; t_{p_1} = \dots = t_{p_k} \} \right|
  = \frac{2^k - 2 + 2^{k \tau'}}{2^k - 1} \prod^{k}_{i=1} \sigma_{p_i} \label{eq:45}
\end{equation}
and
\begin{equation}
  \frac{\left| \{a \in ({\bf Z} / n {\bf Z})^{\times} ; t_{p_1} = \dots = t_{p_k} \} \right|}{\left| ({\bf Z} / n {\bf Z})^{\times} \right|}
  = \frac{1}{2^k - 1} \frac{2^k - 2 + 2^{k \tau'}}{2^{\tilde \tau}}. \label{eq:46}
\end{equation}
\end{The}
{\bf (Proof)}
Applying Eq.(\ref{eq:pq}) and Eq.(\ref{eq:pql}), we have
\begin{eqnarray*}
  &&\left| \{a \in ({\bf Z} / n {\bf Z})^{\times} ; t_{p_1} = \dots = t_{p_k} \} \right| \\
  &&= \left| \bigcup^{\tau'}_{l=0} \{a \in ({\bf Z} / n {\bf Z})^{\times} ; t_{p_1} = \dots = t_{p_k} = l \} \right| \\
  &&= \sum^{\tau'}_{l=0} \left| \{a \in ({\bf Z} / n {\bf Z})^{\times} ; t_{p_1} = \dots = t_{p_k} = l \} \right| \\
  &&= \left| \{a \in ({\bf Z} / n {\bf Z})^{\times} ; t_{p_1} = \dots = t_{p_k} = 0 \} \right|
  + \sum^{\tau'}_{l=1} \left| \{a \in ({\bf Z} / n {\bf Z})^{\times} ; t_{p_1} = \dots = t_{p_k} = l \} \right| \\
  &&= \prod^{k}_{i=1} \sigma_{p_i} + \sum^{\tau'}_{l=1} \left( 2^{k(l-1)} \prod^{k}_{i=1} \sigma_{p_i} \right) \\
  &&= \frac{2^k - 2 + 2^{k \tau'}}{2^k - 1} \prod^{k}_{i=1} \sigma_{p_i}
\end{eqnarray*}
which implies Eq.(\ref{eq:45}).
In addition, since we have
\begin{equation}
  \left| ({\bf Z} / n {\bf Z})^{\times} \right| 
  = \left| ({\bf Z} / p_1 {\bf Z})^{\times} \right| \cdots \left| ({\bf Z} / p_k {\bf Z})^{\times} \right|
  = 2^{\tilde \tau} \prod^{k}_{i=1} \sigma_{p_i},
\end{equation}
we obtain Eq.(\ref{eq:46}) from Eq.(\ref{eq:45}). 

\hfill \qed

\section{A precise estimation of the comutational complexity}

\begin{Lem}
For $n = {p_1}^{e_1} \dots {p_k}^{e_k} $, with prime numbers $p_i,(i=1,\cdots,k)$, we set 
$p_i - 1 = 2^{\tau_i} \sigma_i$ with odd numbers $\sigma_i$,  $\tau_i \geq 1$, $\tau' = \min(\tau_1, \dots, \tau_k)$ and 
$\tilde \tau = \sum^k_{i=1} \tau_i$.
Then we have 
\begin{equation}
  P(A_e \cap A_f \mid A_a \cap A_r) = 1 - \frac{1}{2^k - 1} \frac{2^k - 2 + 2^{k \tau'}}{2^{\tilde \tau}}. \label{eq:51}
\end{equation}
\end{Lem}
{\bf (Proof)}
We pay attention on the properties of the step $5^{\circ}$ and $6^{\circ}$ in the Shor's factoring algorithm presented in section 2 so that we have
\begin{eqnarray*}
  P(A_e \cap A_f \mid A_a \cap A_r)
  &&= P \left( \{ r:even \} \cap \{ a^{r/2} \neq \pm 1 \pmod n \} \Big| A_a \cap A_r \right) \\
  &&= P \left( \{ r:even \} \cap \{ a^{r/2} \neq -1 \pmod n \} \Big| A_a \cap A_r \right) \\
  &&= 1 - P \left( \{ r:odd \} \cup \{ a^{r/2} = -1 \pmod n \} \Big| A_a \cap A_r \right).
\end{eqnarray*}
Here we denote the order of $a$ with respect to $\bmod\,\, n$ and $\bmod\,\, p_i$ by $r = 2^t s$ and $r_i = 2^{t_i} s_i$, respectively.
Where $t \geq 1, t_i \geq 1$ and $s, s_i$ are odd numbers for $i=1,\cdots,k$. Then the probability $P(A_e \cap A_f \mid A_a \cap A_r)$ is rewitten by
\begin{eqnarray*}
  &&P(A_e \cap A_f \mid A_a \cap A_r) \\
  &&= 1 - P \left( \{ r:odd \} \cup \left( \bigcap^k_{i=1} \{ a^{r/2} = -1 \pmod {p_i} \} \right) \Bigg| A_a \cap A_r \right) \\
  &&= 1 - P \left( \{ t_1 = \dots = t_k = 0 \} \cup \left( \bigcap^k_{i=1} \{ t_i = t \} \right) \Bigg| A_a \cap A_r \right) \\
  &&= 1 - P \left( t_1 = \dots = t_k \mid A_a \cap A_r \right).
\end{eqnarray*}
Thus we have Eq.(\ref{eq:51}) from Eq.(\ref{eq:46}). 
\hfill \qed

\begin{The}
For $n = {p_1}^{e_1} \dots {p_k}^{e_k} $, with prime numbers $p_i,(i=1,\cdots,k)$, we set 
$p_i - 1 = 2^{\tau_i} \sigma_i$ with odd numbers $\sigma_i$,  $\tau_i \geq 1$, $\tau' = \min(\tau_1, \dots, \tau_k)$ and 
$\tilde \tau = \sum^k_{i=1} \tau_i$.
Then we have
\begin{equation}
  P_S \geq \left( 1 - \frac{1}{2^k - 1} \frac{2^k - 2 + 2^{k \tau'}}{2^{\tilde \tau}} \right) \frac{\alpha \beta}{\left( \log_2 n \right)^2} \label{eq:52}
\end{equation}
and for any $\varepsilon >0$,
\begin{equation}
N \geq \frac{\log(1/\varepsilon)}{\alpha \beta \left( 1 - \frac{1}{2^k - 1} \frac{2^k - 2 + 2^{k \tau'}}{2^{\tilde \tau}} \right) } \left( \log_2 n \right)^2,
\end{equation}
where $\alpha$ and $\beta$ does not depend on $n$.
\end{The}

{\bf (Proof)}
Applying Eq.(\ref{eq:51}) to Eq.(\ref{eq:PS}) and Eq.(\ref{eq:N}), we have the present theorem. 

\hfill \qed

\section{Comparison of two estimations}
As for Eq.(\ref{eq:PAef}) and Eq.(\ref{eq:51}), we have the following relation.
\begin{Prop} 
For $n = {p_1}^{e_1} \dots {p_k}^{e_k} $, with prime numbers $p_i,(i=1,\cdots,k)$, we set 
$p_i - 1 = 2^{\tau_i} \sigma_i$ with odd numbers $\sigma_i$,  $\tau_i \geq 1$, $\tau' = \min(\tau_1, \dots, \tau_k)$ and 
$\tilde \tau = \sum^k_{i=1} \tau_i$.
Then we have 
\begin{equation}
 1 - \frac{1}{2^k - 1} \frac{2^k - 2 + 2^{k \tau'}}{2^{\tilde \tau}}
  \geq 1 - \frac{1}{2^{k-1}}. \label{eq:61}
\end{equation}
The equality holds when $\tau_1 = \cdots = \tau_k = 1$.
\end{Prop}
{\bf (Prrof)}
Since 
\begin{eqnarray*}
  &&\frac{1}{2^{k-1}} - \frac{1}{2^k - 1} \frac{2^k - 2 - 2^{k \tau'}}{2^{\tilde \tau}} \\
  &&\geq \frac{1}{2^{k-1}} - \frac{1}{2^k - 1} \frac{2^k - 2 - 2^{k \tau'}}{2^{k \tau'}} \\
  &&= \left( \frac{1}{2^k} - \frac{1}{2^{k \tau'}} \right) \left( 1 - \frac{1}{2^k - 1} \right) \\
  &&\geq 0
\end{eqnarray*}
we have inequality. If $\tau_1 = \cdots = \tau_k = 1$, then we easily find the equality holds.
\hfill \qed

Note that Shor's original estimation gives the greatest lower bound of the probability
 $P(A_e \cap A_f \mid A_a \cap A_r) = 1 - \frac{1}{2^k - 1} \frac{2^k - 2 + 2^{k \tau'}}{2^{\tilde \tau}}$.

From now on, we consider the simple case $n=pq$, where $p$ and $q$ are prime numbers. We also set
$p-1 = 2^{\tau_p}\sigma_p$ and $q-1 = 2^{\tau_q}\sigma_q$ where $\tau_p, \tau _q \geq 1$ and $\sigma_p$ and $\sigma_q$ are odd numbers.
By the Shor's original estimation, the probability is given by
$$
  P(A_e \cap A_f \mid A_a \cap A_r) \geq \frac{1}{2}.
$$ 
Moreover, by our precise estimation Eq.(\ref{eq:51}), the probability is given by
$$
  P(A_e \cap A_f \mid A_a \cap A_r) = 1 - \frac{1}{3} \frac{2 + 2^{2 \min(\tau_p, \tau_q)}}{2^{\tau_p + \tau_q}}.
$$

\begin{figure}
  \begin{center}
    \includegraphics[width=12cm, height=9cm]{prob.bmp}
  \end{center}
  \caption{The relation among $\tau_p$, $\tau_q$ and the probability $P(A_e \cap A_f \mid A_a \cap A_r)$.}
  \label{fig:prob}
\end{figure}%

The figure 1 presents the relation among the probability $P(A_e \cap A_f \mid A_a \cap A_r) $, $\tau_p$ and $\tau_q$.
From this figure, we find that the probability $P(A_e \cap A_f \mid A_a \cap A_r) $ takes a minimum value $1/2$ when $\tau_p=\tau_q=1$.
In general, the probability $P(A_e \cap A_f \mid A_a \cap A_r) $ takes small values when $\tau_p=\tau_q$ and it is close to $1$ when $\tau_p \neq \tau_q$.
For the only case that $n=pq$, we can estimate the probability that the number $a (< n)$ satisfying $gcd(a,n) = 1$ is obtained.  

\begin{Prop}
We suppose $n=pq$ with two different prime numbers $p$ and $q$. For sufficient large $n$ and any $\varepsilon >0$, we have 
\begin{equation}
  P(A_a) = \frac{\varphi(n)}{n} \geq \frac{1}{2} \label{eq:paa'}.
\end{equation}
\end{Prop}
{\bf (Proof)}
Firstly we consider the case: $n = pq$. By Euler function, we have
$$
  \frac{\varphi(n)}{n} = \left( 1 - \frac{1}{p} \right) \left( 1 - \frac{1}{q} \right)
  \geq \left( 1 - \frac{1}{2} \right) \left( 1 - \frac{1}{q} \right)
  = \frac{1}{2} \left( 1 - \frac{1}{q} \right).
$$
By the similar way for $q$, we have
$$
  \frac{\varphi(n)}{n} \geq \frac{1}{2} \left( 1 - \frac{1}{p} \right),\,\,
  \frac{1}{2} \left( 1 - \frac{1}{q} \right).
$$
Since we take the limit such as  $n \rightarrow \infty \Longleftrightarrow (p \rightarrow \infty$ or $q \rightarrow \infty)$,
we then have for any $\varepsilon$ and sufficient large $n$, 
$$
  \frac{\varphi(n)}{n} > \frac{1}{2} - \varepsilon .
$$
Secondly we consider the case  $n = 2q$, then we have,
$$
  \frac{\varphi(n)}{n} = \frac{1}{2} \left( 1 - \frac{1}{q} \right).
$$
Taking $q \rightarrow \infty$, we have
$$
  \frac{1}{2} \left( 1 - \frac{1}{q} \right) \longrightarrow \frac{1}{2}.
$$
Therefore we have for any positive number $\varepsilon$,
$$
  \left| \left \{ \frac{\varphi(n)}{n} \,\Big|\, \frac{1}{2} - \varepsilon < \frac{\varphi(n)}{n} <  \frac{1}{2} + \varepsilon \right \} \right| = \infty.
$$
Thus we have
$$
  \liminf_{n \to \infty} \frac{\varphi(n)}{n} = \frac{1}{2}
$$
which implies  Eq.(\ref{eq:paa'}) for sufficient large $n$.

\hfill \qed

\begin{Cor}
We suppose $n=pq$ with two different prime numbers $p$ and $q$.  
Also we set $p - 1 = 2^{\tau_p} \sigma_p, \,q - 1 = 2^{\tau_q} \sigma_q$ and $\tau' = \min(\tau_p, \tau_q),$
where $\sigma_p$ and $\sigma_q$ are odd numbers such that $\tau_p \geq 1$ and $\tau_q \geq 1$. Then we have
\begin{equation}
  P_S \geq \frac{\alpha}{2 \log_2 n}
  \left( 1 - \frac{1}{3} \frac{2 + 2^{2 \tau'}}{2^{\tau_p + \tau_q}} \right) .\label{eq:ps"}
\end{equation}
Also we have  for any $\varepsilon >0$, 
\begin{equation}
  N \geq \frac{2 \log(1 / \varepsilon)}{\alpha \left( 1 - \frac{1}{3}
  \frac{2 + 2^{2 \tau'}}{2^{\tau_p + \tau_q}} \right) } \log_2 n .\label{eq:n"}
\end{equation}
\end{Cor}

\vspace{1cm}

\end{document}